\documentclass[lettersize,journal]{IEEEtran}
   \usepackage{graphicx}
\usepackage[linesnumbered,ruled]{algorithm2e}
\usepackage[noend]{algpseudocode}
\usepackage{booktabs}
\usepackage{amssymb}
\usepackage{amsthm}
\usepackage{multirow} 
\usepackage[cmex10]{amsmath}
\usepackage[caption=false,font=normalsize,labelfont=sf,textfont=sf]{subfig}
\usepackage{textcomp}
\usepackage{stfloats}
\usepackage{xcolor}
\usepackage[noadjust]{cite}

\begin{document}
\title{Near-perfect Coverage Manifold Estimation in Cellular Networks via conditional GAN}
     
\author{Washim Uddin Mondal, Veni Goyal, Satish V. Ukkusuri, Goutam Das, Di Wang, Mohamed-Slim Alouini, {\em Fellow, IEEE}, and Vaneet Aggarwal, {\em Senior Member, IEEE} \thanks{W. U. Mondal, V. Goyal, and V. Aggarwal are with the School of Industrial Engineering, Purdue University, West Lafayette, IN 47907 USA. W. U. Mondal, and S. V. Ukkusuri are with the School of Civil Engineering, Purdue University, West Lafayette, IN 47907 USA. G. Das is with the G. S. Sanyal School of Telecommunications, Indian Institute of Technology Kharagpur, India 721302. D. Wang, M.-S. Alouini, and V. Aggarwal are with the Computer, Electrical and Mathematical Sciences and Engineering division, King Abdullah University of Science and Technology (KAUST), Thuwal 23955-6900, Saudi Arabia. Email: \{wmondal, goyal110, vaneet, sukkusur\}@purdue.edu, gdas@gssst.iitkgp.ac.in, \{di.wang, slim.alouini\}@kaust.edu.sa }}
        
\maketitle

\begin{abstract}
    This paper presents a conditional generative adversarial network (cGAN) that translates base station location (BSL) information of any Region-of-Interest (RoI) to location-dependent coverage probability values within a subset of that region, called the region-of-evaluation (RoE). We train our network utilizing the BSL data of India, the USA, Germany, and Brazil. In comparison to the state-of-the-art convolutional neural networks (CNNs), our model improves the prediction error ($L_1$ difference between the coverage manifold generated by the network under consideration and that generated via simulation) by two orders of magnitude. Moreover, the cGAN-generated coverage manifolds appear to be almost visually indistinguishable from the ground truth. 
\end{abstract}


\IEEEpeerreviewmaketitle

\section{Introduction} 
\label{section_Introduction}

Understanding how the topology of a network influences its performance is one of the central questions in communication networks. Topology-to-performance map is not only crucial for assessing the efficacy of existing networks but also important for designing future ones. One can measure the performance of a network by performing Monte Carlo (MC) simulations over a long time horizon. However, depending on the level of detail incorporated into the simulation framework, this process could demand a high execution time. For example, one way to model channels in the fifth generation and beyond (5G$\&$B) networks is via ray tracing methods \cite{he2018design} that require a large computation time to yield high-fidelity output. Another approach to evaluate the network performance is via stochastic geometry (SG) that yields spatially averaged value of a certain performance metric by incorporating some simplifying assumptions. For example, one of the earliest SG-based models \cite{andrews2011tractable} assumes that the base stations (BS) and users in outdoor wireless networks are placed according to two independent Poisson point processes (PPPs). This helps \cite{andrews2011tractable} to derive an analytical expression for computing the average coverage probability (ACP) of the entire network. The model of PPP was later replaced by more realistic $K$-tier PPP \cite{dhillon2012modeling}, repulsion-based models \cite{deng2014ginibre}, cluster-based models \cite{afshang2018poisson}, $\alpha$-stable processes \cite{li2018stochastic}, etc. Although ACP provides aggregated information about a network's performance, unlike simulation, it cannot characterize the performance metric as a function of users' locations. We want to state that, alongside ACP, another metric, called the meta distribution (MD) \cite{haenggi2015meta}, is also popularly used to evaluate a network's performance. MD is defined to be the cumulative density function (cdf) of coverage values in the network. Although MD provides more information than ACP, the information is still spatially aggregated in nature. Hence, it is impossible to obtain location-specific values of the coverage probability solely from the knowledge of the MD. In summary, SG-based models cannot be true replacements for computation -hungry simulation methods. Moreover, recent SG-based models such as \cite{li2018stochastic} require a fairly complex parameter calibration process, and their expression of ACP incorporates complicated complex integrals that itself could be difficult to evaluate. 

It was shown in our previous work \cite{mondal2022deep} that a properly trained convolutional neural network (CNN) can be significantly better at predicting location-specific coverage probability values than SG-based models. Despite this success, CNN-generated values appear to be noticeably different from the ground truth (Fig. \ref{fig_3} provides a comparison). In this paper, we introduce conditional Generative Adversarial Network (cGAN) architecture to bridge this gap. At its core, the task of predicting coverage probability is similar to the task of image translation which cGAN excels at \cite{isola2017image}. In particular, the network's topology can be represented as a binary matrix whose elements indicate either the presence or the absence of a BS while coverage values can be presented by another matrix with elements in $[0, 1]$. Clearly, both of these matrices can be treated as grayscale images and processed by cGAN. We train our proposed network using the BS location data of India, Brazil, the USA, and Germany and demonstrate that the error performance of our model is at least two orders of magnitude better than that of the CNN and SG-based models. We also show that the coverage values generated by our trained cGAN are visually indistinguishable from the ground truth. It clearly establishes that a properly trained NN can be used as an efficient and reliable replica of a simulator which dramatically accelerates the network evaluation and planning process.

\section{System Model}
\label{section_system_model}

We consider a large area $\mathcal{A}$ where the network performance needs to be evaluated. The space $\mathcal{A}$ can be as large as a country depending on the available dataset. For ease of computation, $\mathcal{A}$ is divided into multiple square-shaped region-of-interest (RoIs) of size $L\times L$ where the exact value of $L$ is specified later. Let, $\mathcal{R}\subset \mathcal{A}$ be an arbitrary RoI, and $\{\mathbf{r}_1, \cdots, \mathbf{r}_n\}$ be the locations of $n$ number of base stations (BSs) that are located within $\mathcal{R}$. In this paper, we solely focus on downlink communication and assume that each user is served by its closest BS. Hence, if the user located at $\mathbf{r}\in \mathcal{R}$ is served by the $j$-th BS, $j\in\{1,\cdots, n\}$, then $j=\arg\min_{l\in\{1, \cdots, n\}}|\mathbf{r}-\mathbf{r}_l|$. The (random) channel gain between the user and the  $l$-th BS is denoted as $g_l$. Clearly, the signal-to-interference-plus-noise-ratio (SINR) experienced at $\mathbf{r}$ can be written as follows.
\begin{align}
\label{eq_SINR}
\mathrm{SINR}(\mathbf{r}) = \dfrac{g_j|\mathbf{r}-\mathbf{r}_j|^{-\alpha}}{\sum_{l\in\{1,\cdots, n\}\setminus\{j\}}g_l|\mathbf{r}-\mathbf{r}_l|^{-\alpha}+\sigma^2/P} ,
\end{align}
where $\alpha$ is the pathloss coefficient, $P$ is the transmission power of the BSs, and $\sigma^2$ is the noise variance. We would like to point out that the SINR expression $(\ref{eq_SINR})$ does not include interferences caused by the BSs lying outside of the RoI, $\mathcal{R}$. This is a good approximation if $L$ is sufficiently large and $\mathbf{r}$ is near the center of $\mathcal{R}$,  thereby ensuring that it is far away from out-of-RoI BSs. However, if $\mathbf{r}$ is located near the edges of $\mathcal{R}$, then $(1)$ may no longer be accurate. To account for this edge effect, we evaluate SINR only within a  concentric square subset of the RoI of size $L/2\times L/2$ which we term as the Region-of-Evaluation (RoE). Using $(\ref{eq_SINR})$, we can obtain the coverage probability at $\mathbf{r}$ (located inside an RoE) as follows.
\begin{align}
\label{eq_cov}
\mathrm{Cov}(\mathbf{r}, \gamma) = \mathrm{Pr}(\mathrm{SINR}(\mathbf{r})>\gamma),
\end{align}
where $\gamma$ is a pre-defined SINR threshold. Note that, by varying $\mathbf{r}$ in $(\ref{eq_cov})$, one can obtain the coverage manifold within the RoE by using the BS location information inside the associated RoI. Although the location variables $\{\mathbf{r}, \mathbf{r}_1,\cdots, \mathbf{r}_n\}$ are continuous in nature,  we must discretize them for computational purposes. This is achieved by discretizing each RoI, and RoE into $N\times N$, and $N/2\times N/2$ square grids respectively where the exact value of $N$ is provided later. Clearly, the BS locations inside an RoI can be described by a binary matrix of size $N\times N$ such that the values of its elements denote either the presence or absence of a BS at the specified location. On the other hand, the coverage manifold inside an RoE can be treated as an $N/2\times N/2$ sized matrix each of whose elements lies in $[0, 1]$.

\section{cGAN-based Coverage Manifold Prediction}
\label{section_gan}

It is clear from the above discussion that the task of translating BS locations (inside an RoI) to a coverage manifold (inside an RoE) is similar to the task of image-to-image translation. As stated earlier, we previously employed a convolutional neural network (CNN)-based auto-encoder to accomplish this job \cite{mondal2022deep}. In this paper, we employ conditional GAN (cGAN) and exhibit that it improves the prediction error performance, as compared to that of the CNN, by two orders of magnitude. 

\subsection{Preliminaries of the cGAN}

Let the BS locations within an RoI be denoted as the matrix $\mathbf{x}$, and the simulated coverage manifold within its RoE be the matrix, $\mathbf{y}$. Our discussion in section \ref{section_system_model} points out that there is a function, $f$ such that $\mathbf{y}=f(\mathbf{x})$, i.e., $\mathbf{y}$ can be fully constructed solely using $\mathbf{x}$. Conditional GAN (cGAN) is one of the popular neural network (NN) architectures that tries to mimic the map, $f$ via supervised learning. cGAN is primarily composed of two main components-the generator, $G_{\boldsymbol{\theta}}$, and the discriminator, $D_{\boldsymbol{\phi}}$, where $\boldsymbol{\theta}$ and $\boldsymbol{\phi}$ are associated neural network parameters. The objective of the generator is to generate a matrix, $G_{\boldsymbol{\theta}}(\mathbf{x})$ from a given input matrix, $\mathbf{x}$ such that it is hard for the discriminator to distinguish $G_{\boldsymbol{\theta}}(\mathbf{x})$ from its corresponding real output matrix $\mathbf{y}$. The loss function of cGAN can be written as follows.
\begin{align}
    \begin{split}
    \mathcal{L}_{cGAN}(\boldsymbol{\theta}, \boldsymbol{\phi}) &= \mathbb{E}\left[\log D_{\boldsymbol{\phi}}(\mathbf{x}, \mathbf{y})\right] \\
    &+\mathbb{E}\left[\log(1-D(\mathbf{x}, G_{\boldsymbol{\theta}}(\mathbf{x})))\right],
    \end{split}
\end{align}
where the expectations are computed over the joint distribution of $(\mathbf{x}, \mathbf{y})$, and $D_{\phi}(\mathbf{x}, \mathbf{y})$ defines the probability assigned by the discriminator of $\mathbf{y}$ being the real output of $\mathbf{x}$. Mathematically, the objective function of cGAN be expressed as follows.
\begin{align}
    \begin{split}
        &\min_{\boldsymbol{\theta}} \max_{\boldsymbol{\phi}} \mathcal{L}_{cGAN}(\boldsymbol{\theta}, \boldsymbol{\phi}) + \lambda \mathcal{L}_1(\boldsymbol{\theta}),\\
        \text{where} ~&\mathcal{L}_1(\boldsymbol{\theta}) \triangleq \mathbb{E}\left|\mathbf{y}-G_{\boldsymbol{\theta}}(\mathbf{x})\right|_1.
    \end{split}
\end{align}

\begin{figure}
    \subfloat[\label{fig_1a}]{\includegraphics[width=0.83\linewidth]{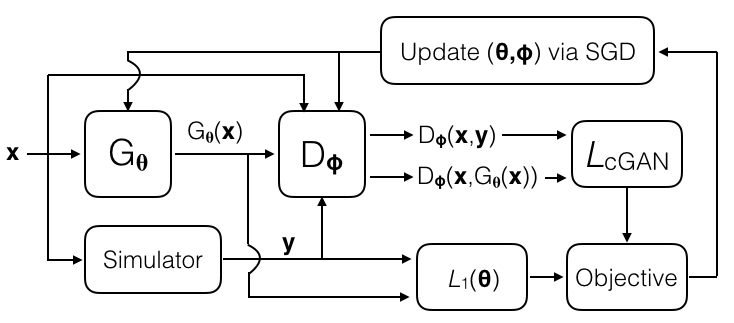}}
    \subfloat[\label{fig_1b}]{\includegraphics[width=0.16\linewidth]{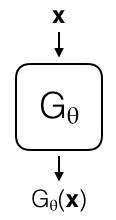}}
    \caption{Working principle of cGAN. Fig. \ref{fig_1a} depicts the training process while Fig. \ref{fig_1b} describes how to generate output from a trained network. The matrix, $\mathbf{x}$ describes the location of BSs in an RoI whereas the matrix, $\mathbf{y}$  defines the simulated coverage manifold within its RoE.}
    \label{fig_1}
\end{figure}

The regularizer $\mathcal{L}_1(\boldsymbol{\theta})$ is added to ensure that, in addition to fooling the discriminator, the generator also generates matrices that are close to the real output. The function $|\cdot|_1$ denotes the $L_1$ norm, and $\lambda>0$ is a regularizer parameter. Fig. \ref{fig_1} portrays the working principle of cGAN.

\subsection{Proposed Architecture}
\label{section_sub_architecture}

We use U-Net \cite{ronneberger2015u} architecture for the generator. Note that the size of the input (BS locations in an RoI) is $N\times N$ whereas that of the output (coverage manifold in an RoE) is $N/2\times N/2$. Thus, both can be used as grayscale images with one channel. We first pass the input through a convolutional layer with stride $2$ followed by a Leaky ReLU activation layer so that its output appears to be of the same size as that of the coverage manifold in an RoE. The resulting matrix is then passed through a series of encoders followed by a series of decoders. Each encoder is composed of a convolutional layer, a Leaky ReLU layer, and a BatchNorm Layer. Convolutional layers are further composed of $512$ convolution filters, each with kernel size $4$, and stride $2$. On the other hand, a decoder is composed of a ReLU module, a deconvolution layer (containing $512$ deconvolution filters with kernels of size $4$, and stride $2$), and a BatchNorm layer. In the last decoder, BatchNorm operation is replaced by an activation function, $g$ where $g(x) = (1+\tanh(x))/2$ so that the elements of the output matrix lie in $[0, 1]$. Skip connections are added between pairwise encoder and decoder layers by concatenating the input of the encoder to the output of the decoder. This helps retain information at various depths/scales and circumvents the vanishing gradient problem.

\begin{figure}
    \centering
    \subfloat[\label{subfig_generator}]{\includegraphics[width=0.9\linewidth]{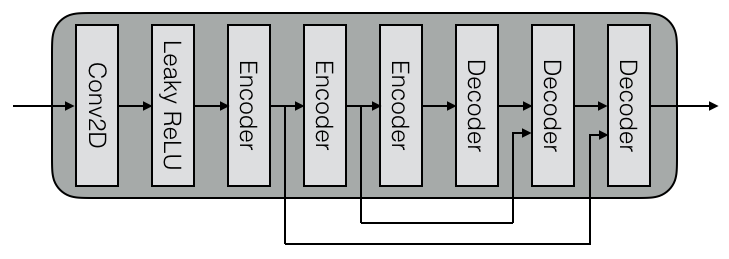}} \vspace{0.1cm}
    \subfloat[\label{subfig_discriminator}]{\includegraphics[width=0.54\linewidth]{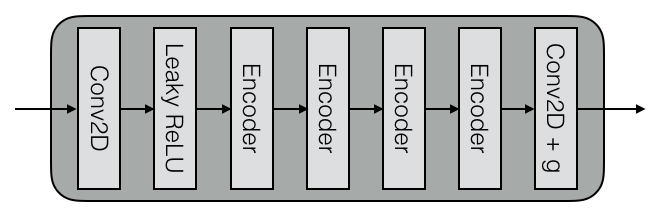}}
    \subfloat[\label{subfig_encoderdecoder}]{\includegraphics[width=0.45\linewidth]{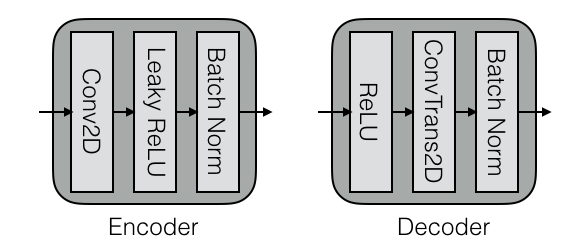}}
    \caption{Architectural details of cGAN. Fig. \ref{subfig_generator}, \ref{subfig_discriminator} describe the structure of the generator and discriminator respectively while Fig. \ref{subfig_encoderdecoder} depicts the structure of encoder and decoder blocks.}
    \label{fig_architecture}
\end{figure}

The discriminator is modeled using a PatchGAN \cite{isola2017image} architecture. Its input is sequentially passed through a convolution block, a Leaky ReLU function, and a series of encoder blocks. Each convolutional layer comprises $512$ filters with kernels of size $4$, and stride $2$.  The primary characteristic of a PatchGAN discriminator is that it only attempts to identify whether a part of the given input is real. In contrast, an ordinary discriminator attempts to do the same for the entire input. In line with this philosophy, the output of the last encoder is passed through a convolution layer followed by the activation, $g$ defined earlier. This yields decision probabilities of the discriminator network corresponding to each receptive field of the convolution layer. We then average all these probabilities to produce the decision probability of the entire input. Fig. \ref{fig_architecture}  depicts each component of the cGAN architecture.

\subsection{Training and Testing Procedures}

In order to train the proposed cGAN architecture, we utilize BS location datasets\footnote{Collected from the website  www.opencellid.org} of four different countries, namely India, the USA, Germany, and Brazil. The coordinates of the BSs are projected onto a flat geometry via the following approach. Let the spherical coordinates of two closely located BSs be $(\Theta, \Phi)$, and $(\Theta+\Delta \Theta, \Phi+\Delta \Phi)$. In the flat geometry, their difference in coordinates maps to $(R\Delta \Theta, R\cos(\Theta)\Delta \Phi)$  where $R$ indicates the average radius of the Earth. Now we partition each country into multiple RoIs of size $L\times L$. Next, we discard those RoIs that contain either less than $B_{\min}$ or more than $B_{\max}$ number of BSs where the values of $B_{\min}$ and $B_{\max}$ are specified later. The reason behind imposing the above thresholds is to remove the outlier cases. In the remaining pool of RoIs, we randomly select $70\%$ for training the proposed cGAN while the rest $30\%$ are used for testing its performance. We use stochastic gradient descent (SGD) with Adam optimizer to train the network.

The performance of cGAN is assessed via $L_1$ error. Specifically, if $\boldsymbol{\theta}^*$ is the trained parameter of the generator,  $\{\mathbf{x}_k\}_{k=1}^K$ are the locations of BSs inside RoIs that are used in the testing phase and $\{\mathbf{y}_k\}_{k=1}^K$ denote their associated simulated coverage manifolds, then the $L_1$ error is defined as,
\begin{align}
    \label{eq_error_cGAN}
    \mathrm{error}(\mathrm{cGAN}) = \dfrac{1}{K}\sum_{k=1}^K \left|\mathbf{y}_k-G_{\boldsymbol{\theta}^*}(\mathbf{x}_k)\right|_1.
\end{align} 

To demonstrate the superiority of our method, we compare the $L_1$ error performance of cGAN, with that of a trained CNN model \cite{mondal2022deep}. We also compare the performance of cGAN to that of a PPP and the best-fitted SG model (explained below). Note that all SG-based models estimate only the (spatially) averaged coverage probability (ACP) and not its location-specific value. For example, the ACP estimated by a PPP is given by Theorem 1 in \cite{andrews2011tractable} which is only dependent on the BS density. Therefore, if $\hat{\lambda}(\mathbf{x}_k)$ denotes the empirical BS density of the RoI, $\mathbf{x}_k$, and $\mathrm{ACP}_{\mathrm{PPP}}(\hat{\lambda}(\mathbf{x}_k))$ defines its ACP computed by the PPP model, then the associated estimated coverage manifold can be written as $\mathrm{ACP}_{\mathrm{PPP}}(\hat{\lambda}(\mathbf{x}_k))\mathbf{1}_{{N}/{2}\times {N}/{2}}$ where $\mathbf{1}_{P\times P}$ is a square matrix of size $P$ with all entries equal to one. The error performance of the PPP can now be calculated following a similar equation to $(\ref{eq_error_cGAN})$. On the other hand, the ACP of the RoI, $\mathbf{x}_k$, as estimated by the best-fitted SG (BFSG) model is given by $\mathrm{avg}(\mathbf{y}_k)$ where $\mathbf{y}_k$, as stated before, is the simulated coverage manifold of $\mathbf{x}_k$, and $\mathrm{avg}(\cdot)$ denotes the average function. Hence, the associated coverage manifold yielded by the BFSG is $\mathrm{avg}(\mathbf{y}_k)\mathbf{1}_{N/2\times N/2}$, and its performance can be quantified similar to $(\ref{eq_error_cGAN})$. Note that the BFSG model provides an upper bound to the performances of all SG-based models as they attempt to estimate $\mathrm{avg}(\mathbf{y}_k)$.

\section{Numerical Results}
\label{section_results}
\textit{Parameters:} For the USA, and Germany, we take the size of an RoI to be $L=5$ km because of their dense BS deployment while for India, and Brazil, we use $L=10$ km. For all datasets, the discretization parameter is chosen as $N=256$ which leads to $19.53$ m of resolution for the USA, and Germany and $39.06$ m of resolution for India, and Brazil. The pathloss coefficient is taken as $\alpha=4$, and the BS-user channels gains are assumed to be exponentially distributed with mean $1$. The minimum and maximum thresholds for the number of BS in an RoI are taken as $B_{\min}=20, B_{\max} = 400$ respectively. We assign $\sigma^2/P=0$ as the selected RoIs are interference limited. Finally, to project the BS coordinates onto a flat geometry, the Earth is presumed to be a perfect sphere with $R=6371$ km radius. 

\begin{table}[t!]
    \centering
    \subfloat[India]{
	\centering
	\begin{tabular}{|c|c|c|c|c|c|}
		\hline
		$\gamma$ & $0$ dB & $5$ dB & $10$ dB & $15$ dB & $20$ dB\\
		\hline
		PPP & 0.2750 & 0.3700 & 0.3157  & 0.2233  & 0.1432 \\
		\hline
		BFSG & 0.2312 & 0.2944 & 0.2274 & 0.1561 & 0.0995 \\
		\hline
		CNN & 0.1977 & 0.1551 & 0.0940 & 0.0530 & 0.0331 \\
		\hline
		cGAN & 0.0068 & 0.0012 & 0.0011 & 0.0007 & 0.0004 \\
		\hline
	\end{tabular}
    }
    
    \subfloat[USA]{
	\centering
	\begin{tabular}[t]{|c|c|c|c|c|c|}
		\hline
		$\gamma$ & $0$ dB & $5$ dB & $10$ dB & $15$ dB & $20$ dB\\
		\hline
		PPP & 0.2155 & 0.3160  & 0.3076 & 0.2376 & 0.1602 \\
		\hline
		BFSG & 0.1454 & 0.2962 & 0.2825 & 0.2152 & 0.1449 \\
		\hline
		CNN & 0.1279 & 0.2325 & 0.1527 & 0.0891 & 0.0560 \\
		\hline
		cGAN & 0.0065 & 0.0019 & 0.0014 & 0.001 & 0.0006 \\
		\hline
	\end{tabular}
    }	
    
    \subfloat[Germany]{	
	\centering
	\begin{tabular}[t]{|c|c|c|c|c|c|}
		\hline
		$\gamma$ & $0$ dB & $5$ dB & $10$ dB & $15$ dB  & $20$ dB \\
		\hline
		PPP & 0.2034 & 0.2974 & 0.3056 & 0.2460 & 0.1698 \\
		\hline
		BFSG & 0.1512 & 0.2790 & 0.2989 & 0.2412 & 0.1676 \\
		\hline
		CNN & 0.1285 & 0.2479 & 0.1662 & 0.0970 & 0.0616 \\
		\hline
		cGAN & 0.0014 & 0.0017 & 0.0015 & 0.0011 & 0.0007 \\
		\hline
	\end{tabular}
    }	
    
    \subfloat[Brazil]{
	\centering
	\begin{tabular}[t]{|c|c|c|c|c|c|}
		\hline
		$\gamma$ & $0$ dB & $5$ dB & $10$ dB & $15$ dB & $20$ dB \\
		\hline
		PPP & 0.4266 & 0.4312 & 0.3268 & 0.2165 & 0.1338 \\
		\hline
		BFSG & 0.3039 & 0.2589 & 0.1786 & 0.1158 & 0.0716 \\
		\hline
		CNN & 0.1827 & 0.1420 &  0.0898 & 0.0523 & 0.0331 \\
		\hline
		cGAN & 0.0021 & 0.0018 & 0.001 & 0.0006 & 0.0004 \\
		\hline
	\end{tabular}
    }
    \vspace{0.1cm}
    \caption{Comparison of the error performance of cGAN with that of the PPP, the best fitted stochastic geometric (SG) model, and CNN in four different countries for various values of $\gamma$.}
    \label{table_1}
\end{table}

\begin{figure}
    \centering	
    \subfloat[CNN\label{subfig_cnn_india_105}]{
	\includegraphics[width=0.15\textwidth]{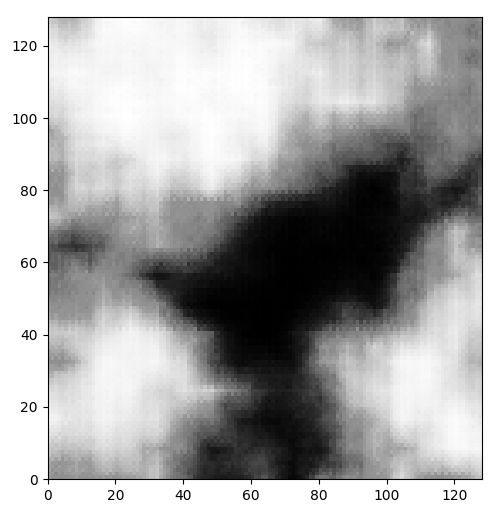}
    }
    \subfloat[cGAN\label{subfig_cgan_india_105}]{
	\includegraphics[width=0.15\textwidth]{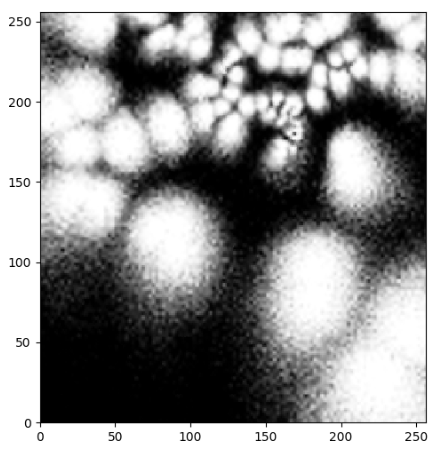}
    }
    \subfloat[Ground Truth\label{subfig_sim_india_105}]{
	\includegraphics[width=0.17\textwidth]{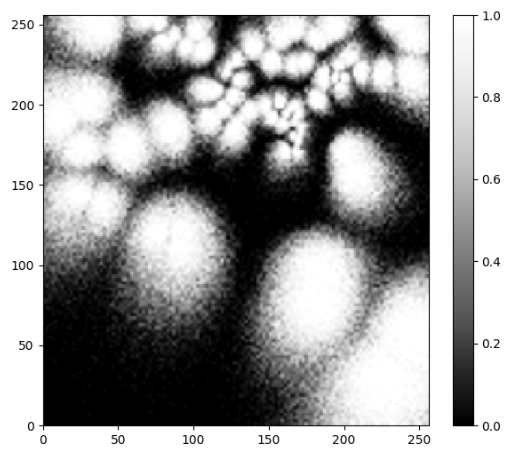}
    }\\
    \subfloat[CNN\label{subfig_cnn_india_125}]{
	\includegraphics[width=0.15\textwidth]{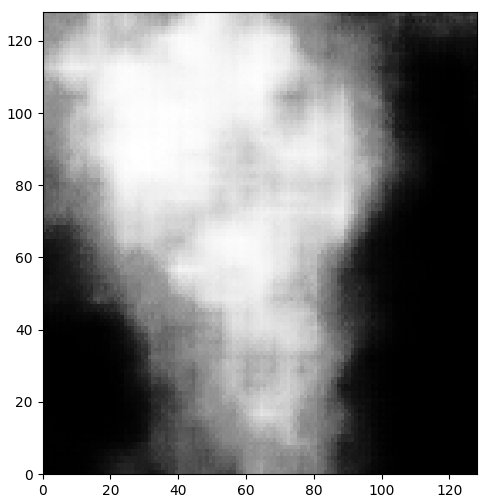}
    }
    \subfloat[cGAN\label{subfig_cgan_india_125}]{
	\includegraphics[width=0.15\textwidth]{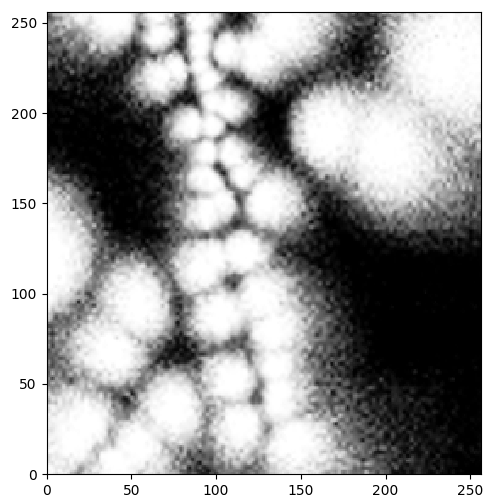}
    }
    \subfloat[Ground Truth\label{subfig_sim_india_125}]{
	\includegraphics[width=0.15\textwidth]{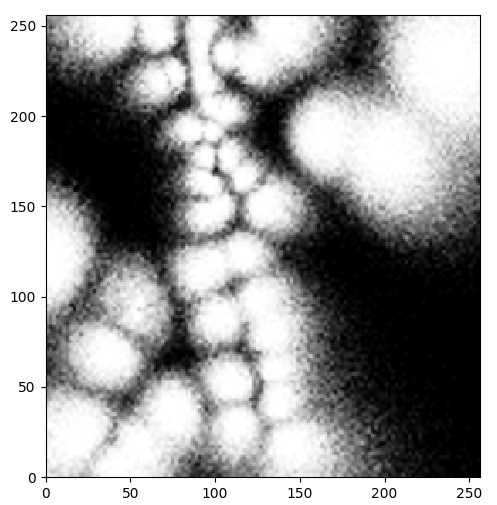}
    }\\
    \subfloat[CNN\label{subfig_cnn_india_170}]{
	\includegraphics[width=0.15\textwidth]{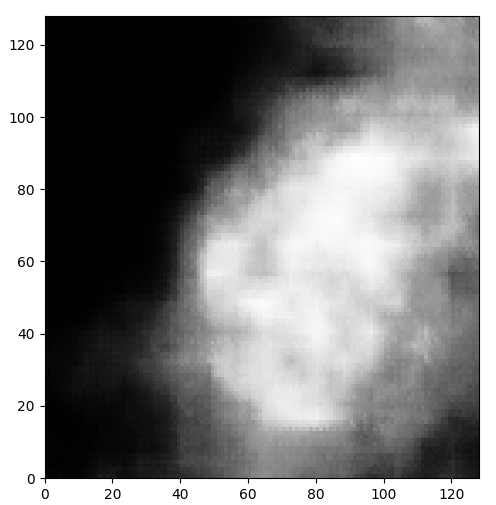}
    }
    \subfloat[cGAN\label{subfig_cgan_india_170}]{
	\includegraphics[width=0.15\textwidth]{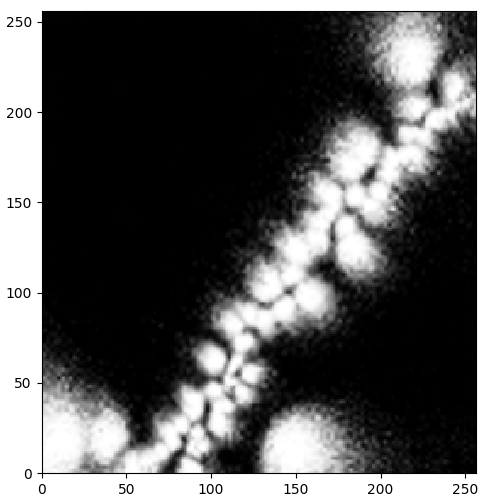}
    }
    \subfloat[Ground Truth\label{subfig_sim_india_170}]{
	\includegraphics[width=0.15\textwidth]{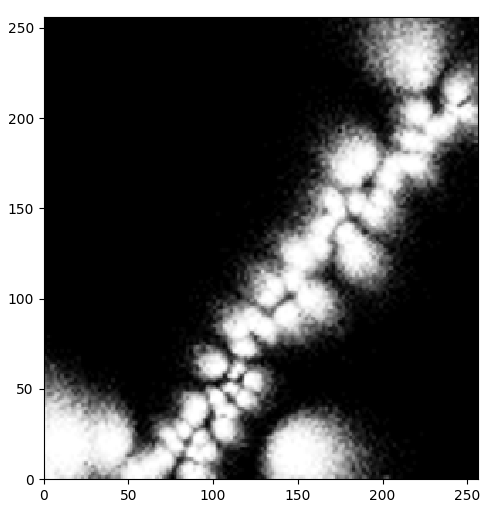}
    }
    \caption{Visual comparison of the coverage manifolds estimated by cGAN and CNN with the ground truth. RoEs of size $128\times 128$ are randomly chosen from the available dataset to generate the manifolds. SINR threshold is $0$ dB.}
    \label{fig_3}
\end{figure}

The main result of our paper is provided in Table \ref{table_1} where we list the error performance of four different approaches, namely cGAN, CNN, BFSG, and PPP for four different countries and a wide range of values of the SINR threshold, $\gamma$. Observe that in all of these cases, the error corresponding to cGAN is lower than that of other methods by at least two orders of magnitude. On the other hand, although the error performance of the CNN is better than that of PPP and BFSG, they are of the same order of magnitude. Another interesting trend observed in this result is that the performances of all models improve with an increase in $\gamma$. This may be because a high threshold shortens the typical range of coverage probabilities, thus making prediction easier. In Fig. \ref{fig_3}, we visually compare the ground truth manifold with that generated by the cGAN and the CNN. It is noticed that the output of cGAN is visually indistinguishable from the ground truth and the output of the CNN is remarkably distinct. Though we report results here for Rayleigh distribution only for lack of space, we note similar results for other channel models such as Nakagami, probabilistic blocking, etc. cGAN architecture takes around 4 hours for training with a 64 GB RAM, 3.0 GHz processor, and an NVIDIA A100-SXM4-40GB GPU whereas an efficient implementation of CNN-AE takes around 3 hours to train. Since NNs are trained via an offline process, a slight increase in training time has little impact on network planning applications. After training, both CNN-AE and cGAN provide similar execution times (1.93 sec and 2.07 sec, respectively). However, the latter NN offers a significant improvement in the prediction accuracy as shown in Table \ref{table_1}.

\section{Conclusion}
\label{section_conclusion}

We propose a cGAN architecture that uses BSL data to yield coverage manifolds. In comparison to the state-of-the-art NNs, our model improves the error by two orders of magnitude. This paper assumes an LTE technology with omnidirectional transmission of BSs. However, extending this study for directional coordinate multi-point (CoMP) transmissions in emerging 5G and O-RAN technologies is an important future consideration.

\bibliography{BibFile.bib}
\bibliographystyle{IEEEtran}

\end{document}